\documentclass[%
 reprint,
 amsmath,amssymb,
aps,
pra,
superscriptaddress,
longbibliography,
]{revtex4-1}

\usepackage{graphicx}
\usepackage{dcolumn}
\usepackage{bm}
\usepackage{hyperref}
\usepackage{color}
\usepackage{tabularx}
\usepackage{float}
\usepackage{upgreek}
\usepackage{nicefrac} 
\usepackage{enumitem}
\usepackage{placeins}
\usepackage{tabularx}
\usepackage{braket}
\usepackage{acronym}
\usepackage[textsize=tiny, backgroundcolor=red!40, linecolor=red]{todonotes}

\newsavebox{\foobox}
\newcommand{\slantbox}[2][0]{\mbox{%
        \sbox{\foobox}{#2}%
        \hskip\wd\foobox
        \pdfsave
        \pdfsetmatrix{1 0 #1 1}%
        \llap{\usebox{\foobox}}%
        \pdfrestore
}}
\newcommand\unslant[2][-.25]{\slantbox[#1]{$#2$}}

\definecolor{rainbow0}{rgb}{0.5, 0.0, 1.0}
\definecolor{rainbow1}{rgb}{0.31960784313725488, 0.27958259259674378, 0.98998021328070696}
\definecolor{rainbow2}{rgb}{0.13921568627450981, 0.53686659764417999, 0.96012164537462819}
\definecolor{rainbow3}{rgb}{0.041176470588235259, 0.75133188955687324, 0.91102264924608833}
\definecolor{rainbow4}{rgb}{0.22941176470588232, 0.91102264924608822, 0.84034407163789271}
\definecolor{rainbow5}{rgb}{0.40980392156862744, 0.98998021328070696, 0.75538273471899375}
\definecolor{rainbow6}{rgb}{0.59019607843137245, 0.98998021328070696, 0.65528385001345368}
\definecolor{rainbow7}{rgb}{0.77058823529411757, 0.91102264924608845, 0.54205335647244945}
\definecolor{rainbow8}{rgb}{0.95882352941176463, 0.75133188955687347, 0.41235631747390367}
\definecolor{rainbow9}{rgb}{1.0, 0.53686659764418021, 0.27958259259674395}
\definecolor{rainbow10}{rgb}{1.0, 0.27958259259674401, 0.14120615182309149}

\begin{document}

\title{Characterization of the T center in $^{28}$Si}

\author{L. Bergeron}
\affiliation{Department of Physics, Simon Fraser University, Burnaby, British Columbia, Canada V5A 1S6}
\author{C. Chartrand}
\affiliation{Department of Physics, Simon Fraser University, Burnaby, British Columbia, Canada V5A 1S6}
\author{A. T. K. Kurkjian}
\affiliation{Department of Physics, Simon Fraser University, Burnaby, British Columbia, Canada V5A 1S6}
\author{K. J. Morse}
\affiliation{Department of Physics, Simon Fraser University, Burnaby, British Columbia, Canada V5A 1S6}

\author{H.~Riemann}
\affiliation{Leibniz-Institut f\"ur Kristallz\"uchtung, 12489 Berlin, Germany}

\author{N.~V.~Abrosimov}
\affiliation{Leibniz-Institut f\"ur Kristallz\"uchtung, 12489 Berlin, Germany}

\author{P. Becker}
\affiliation{Physikalisch-Technische Bundestanstalt Braunschweig, 38116 Braunschweig, Germany}

\author{H.-J. Pohl}
\affiliation{VITCON Projectconsult GmbH, 07743 Jena, Germany}

\author{M. L. W. Thewalt}
\affiliation{Department of Physics, Simon Fraser University, Burnaby, British Columbia, Canada V5A 1S6}

\author{S. Simmons}
\thanks{s.simmons@sfu.ca}
\affiliation{Department of Physics, Simon Fraser University, Burnaby, British Columbia, Canada V5A 1S6}

\date{\today}

\setlength{\skip\footins}{0.5cm}

\begin{abstract}
{Silicon is host to two separate leading quantum technology platforms: integrated silicon photonics as well as long-lived spin qubits. There is an ongoing search for the ideal photon-spin interface able to hybridize these two approaches into a single silicon platform offering substantially expanded capabilities. A number of silicon defects are known to have spin-selective optical transitions, although very few of these are known to be in the highly desirable telecommunications bands, and those that do often do not couple strongly to light. Here we characterize the T center in silicon, a highly stable silicon defect which supports a short-lived bound exciton that upon recombination emits light in the telecommunications O-band. In this first study of T centers in $^{28}$Si, we present the temperature dependence of the zero phonon line, report ensemble zero phonon linewidths as narrow as 33(2) MHz, and elucidate the excited state spectrum of the bound exciton. Magneto-photoluminescence, in conjunction with magnetic resonance, is used to observe twelve distinct orientational subsets of the T center, which are independently addressable due to the anisotropic g factor of the bound exciton's hole spin. The T center is thus a promising contender for the hybridization of silicon's two leading quantum technology platforms. }
\end{abstract}

\maketitle

There is a global search underway to identify a photon-spin interface which is ideally suited to the needs of networked quantum technologies. 
In the sister paper to this work \cite{SisterPRL}, the silicon T center has been newly identified as a leading candidate in this search due to its native operation in the telecommunication O-band, its long-lived spins, and the commercial dominance of its host material. 
Prior to this work, relatively little was known about the relevant quantum properties of the silicon T center. 
Here we report upon a variety of isotopic, luminescence, magneto-optical, thermal, bound exciton excited state, and spin properties of the T center which further support its position as a photon-spin frontrunner in the technological race for a global quantum platform.

Luminescent silicon defects have been studied for decades, with the vast majority of this work taking place before the advent of quantum technology hardware development in the late 1990s. 
Here we revisit the T center, a member of the class of luminescent defects called radiation damage centers. 
These centers can be formed by electron, neutron, or ion radiation damage followed by an annealing treatment, although some radiation damage centers, including the T center, can be created with heat treatment alone \cite{Safonov1996, Lightowlers1994hydrogen}.

Silicon radiation damage centers have themselves been the subject of a number of recent studies \cite{Chartrand2018, Buckley2017, Beaufils2018, Redjem2020} due to their bright luminescence, sub-microsecond luminescence lifetimes, near radiatively-limited optical linewidths in $^{28}$Si, and wavelengths in the telecommunications bands. 
However, despite early reports that the G center in particular is connected to an optically detected magnetic resonance (ODMR) signal \cite{Lee1982}, the dominant zero phonon line (ZPL) transitions of the G, as well as the C and W radiation damage centers, have each been shown to be singlet-to-singlet transitions with no ground state unpaired electron or hole spins in which to hold quantum information. 
The little-studied T center, by contrast, was thought to have a \mbox{spin-1/2} to \mbox{spin-1/2} optical transition directly in the telecommunications O-band, although the composition of the ground state was in dispute \cite{Irion1985, Safonov1996}. 
Encouraged by this, and the knowledge that many defects in isotopically purified $^{28}$Si boast exceptionally long electron \cite{Steger2012, Wolfowicz2012} and nuclear \cite{Saeedi2013} spin lifetimes, the goal of this work and its sister publication \cite{SisterPRL} is to assess the T center for its potential as a silicon telecom photon-spin interface.

This paper is organized as follows. 
In Section~\ref{sec:Review} we begin with a brief introduction and review of what has been reported in the literature about the T center. 
In Section~\ref{sec:Luminescence spectra} we present photoluminescence (PL) results and previously unobserved isotopic shifts. 
These results add credibility to the existing atomic model of the center. 
In Section~\ref{sec:ZPL Temperature Dependence} we apply photoluminescence excitation (PLE) spectroscopy to reveal an ensemble ZPL full-width half-maximum (FWHM) linewidth as narrow as 33(2)\,MHz, and report upon the temperature dependence of the ZPL, whose linewidth, position and amplitude all vary with temperature over the range of 1.2\,K\,--\,4.2\,K in a predictable manner. 
In Section~\ref{sec:Bound Exciton Excited States} we use PLE to identify the energies of many of the T center's bound exciton (BE) excited states. 
In Sections \ref{sec:Magneto-photoluminescence}--\ref{sec:Pulsed spin resonance} we study T centers under an applied magnetic field. 
Section \ref{sec:Magneto-photoluminescence} shows magneto-PL, which reveals that the T center has a number of different orientations which become inequivalent when a magnetic field is applied in an arbitrary direction. 
In Section~\ref{sec:Continuous-wave spin resonance} we reveal a hyperfine interaction with the hydrogen nuclear spin, and explore the combination of continuous-wave spin resonance and resonant optical excitation for electron and nuclear spin state hyperpolarization and readout.
In Section~\ref{sec:Pulsed spin resonance} we study a specific orientational subset to measure spin qubit characteristics using pulsed magnetic resonance.
We conclude with a discussion of prospects and future work. 
The appendices include technical details on the sample preparation methods and the experimental apparatuses used in this work.

\section{Review}
\label{sec:Review}

In the 1970s, luminescent defects in silicon, and in particular sharp spectral features created as a result of irradiation and/or heat treatment, were often labeled with a letter. 
A number of these so-called radiation damage centers have been studied extensively \cite{Davies1989, Davies2006}. 
Most centers in this category have ZPL transitions which are unsplit by magnetic fields, with no unpaired electron spins in their unexcited states, limiting their usefulness as photon-spin interfaces. 
The T \cite{Minaev1981, Irion1985, Henry1991, Safonov1994, Lightowlers1994hydrogen, Lightowlers1994luminescence, Safonov1996, Leary1998, Safonov1999, Schmidt2000, Hayama2004, Davies2006}, I \cite{Henry1991, Lightowlers1994hydrogen, Gower1997, Safonov1999}, and M \cite{Lightowlers1994luminescence, Safonov1994, Lightowlers1997, Safonov1999, Schmidt2000} centers are notable exceptions to this general trend.

\textit{Level structure} -- In 1981, Minaev and Mudryi \cite{Minaev1981} discovered that the ZPL of the `T line' photoluminescence feature near 935\,meV was in fact a doublet. 
In 1985, Irion et al. \cite{Irion1985} confirmed that the 1.8 meV--split doublet (later estimated in Ref.~\cite{Safonov1996} to be 1.75\,meV) is a result of two states in the same defect, which we will refer to as TX$_0$ and TX$_1$. 
In the same work, Irion et al. presented the nonlinear and orientation-dependent stress dependence of the T center's TX$_0$ and TX$_1$ ZPL optical transitions, revealing a rhombic-I (C$_\textrm{2v}$) defect symmetry. 
Later studies \cite{Safonov1996} revealed even more splitting under stress, indicating that the T center is of monoclinic I (C$_\textrm{1h}$) symmetry.

\begin{figure}
    \includegraphics[width=\linewidth]{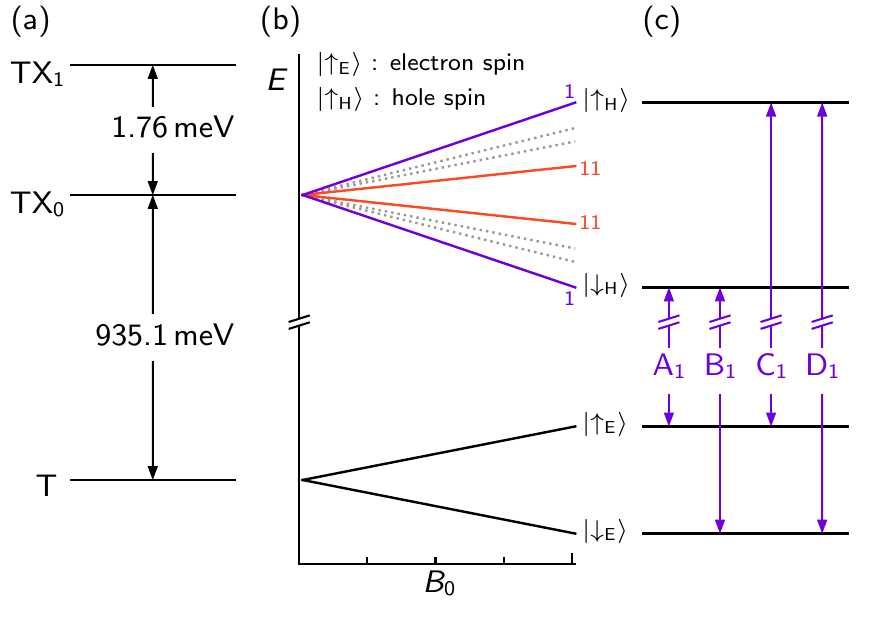}
    \caption{(a) Schematic energy level diagram at zero field, not to scale, showing the ground T level, and the bound-exciton TX$_0$ and TX$_1$ states. (b) Energy levels for orientational subset 1 in purple, not to scale, as a function of magnetic field amplitude $| B_0 |$. Grey and red lines represent example TX$_0$ states for orientational subsets 2 to 11, with different splittings for a given B$_0$ field orientation caused by the anisotropic g factor of the TX$_0$ hole spin. (c) Optical transitions accessible for orientational subset 1, relevant to the discussion in Sections \ref{sec:Magneto-photoluminescence}--\ref{sec:Pulsed spin resonance}. In this work, nuclear spin-selective transitions are not resolvable optically. The four available optical transitions for subset 1 are labeled A$_1$, B$_1$, C$_1$, D$_1$, which is consistent with the labels used throughout this manuscript.
    \label{fig:diagram}
    }
    \end{figure}

Based upon early magnetic field dependence studies \cite{Irion1985}, Irion et al. concluded that the TX$_0$ ZPL consists of a transition between a level possessing a highly anisotropic \mbox{spin-1/2} particle and a level possessing a highly isotropic \mbox{spin-1/2} particle. 
The isotropic and anisotropic spin g factors were determined in Ref. \cite{Irion1985}. 
Irion et al proposed, erroneously, that an anisotropic hole spin and an isotropic electron spin both reside in the bound exciton state, which this work disproves. 
The magnetic resonance results in Section \ref{sec:Continuous-wave spin resonance} are consistent with the later model proposed by Safonov et al. \cite{Safonov1996}. 
In this model the T ground state has an unpaired electron, and the TX state includes an additional bound exciton. The two bound electrons in the TX state pair into an s\,=\,0 singlet state and the unpaired j\,=\,3/2 hole spin state is split by the reduced symmetry of the defect, which can be modelled as an internal stress, into two doublets, TX$_0$ and TX$_1$. 
The unpaired hole spin determines the magnetic splitting of the TX$_0$ state. 
The total binding energy of the electron-hole pair is approximately 235\,meV, whereas the PL intensity decays with an activation energy of approximately $\sim$\,22\,meV \cite{Irion1985} or $\sim$\,32\,meV \cite{Safonov1996}, which reflects the binding energy of the hole to the negatively-charged T$^-$ center. 
A T center level diagram, which summarizes these earlier findings, is shown in Fig.~\ref{fig:diagram}(a and b). 

\textit{Atomic composition and formation} -- Studies in 1985 \cite{Irion1985}, and then later in 1996 \cite{Safonov1996}, identified the presence of at least one carbon atom and one hydrogen atom in the T center, by observing shifts and splittings of the T center TX$_0$ ZPL with the incorporation of $^{13}$C and deuterium in the studied samples.
Furthermore, a fourfold splitting observed in a local vibrational mode (LVM) seen in PL revealed that there are two inequivalent carbon atoms present in the defect \cite{Safonov1996}. 
There was early evidence in 1981 that oxygen is not involved in the formation of the T center (as opposed to the I center) \cite{Minaev1981}. 
The resemblance of the T center with Al1 \cite{Irion1988} and Ga1 \cite{Thonke1985} defects, with regard to LVM shifts, carbon isotope shift, and alleged symmetry, led to early speculation that boron was involved in the atomic structure of the T center \cite{Irion1985, Thonke1985} -- there is however no direct evidence supporting this conjecture.

Ab-initio cluster calculations \cite{Safonov1996, Leary1998}, drawing upon a detailed analysis of the PL sideband LVM data first presented in \cite{Irion1985} and expanded upon in \cite{Safonov1996}, predicted an atomic structure for the T center. 
In this model, two carbon atoms are directly bonded and together share the substitutional site of a silicon atom; one of these carbon atoms is terminated with a hydrogen atom, leaving an unpaired electron dangling bond on the other carbon atom \cite{Safonov1996}. 
The paramagnetic nature of the T, I, and M centers in their neutral ground states are consistent with the presence of one hydrogen atom in each of these defects \cite{Safonov1999}.

\textit{T center formation} -- While it has been suggested that M centers are involved in the formation of T centers \cite{Irion1985, Safonov1999}, it has been claimed \cite{Lightowlers1994hydrogen} that M centers have never been observed in non-irradiated material, as opposed to T centers. 
From ab-initio studies \cite{Safonov1996, Leary1998}, the suggested T center formation mechanism is the capture of an interstitial C-H pair by a substitutional carbon atom \cite{Safonov1996, Leary1998, Davies2006}. 
T centers have been observed both in Czochralski (CZ) grown and float zone (FZ) grown silicon \cite{Safonov1996}, and the major step common to all studies is the need to apply a heat treatment with temperature between 350\,$^\circ$C and 600\,$^\circ$C. 
Hydrogen can be already present in the silicon, or introduced by water vapour or gaseous hydrogen during the thermal anneal \cite{Lightowlers1994hydrogen}. 
It has been observed that excess hydrogen in the silicon sample passivates the T center \cite{Leary1998}, rendering it optically inactive.

\section{Luminescence spectra}
\label{sec:Luminescence spectra}

In the PL experiments using nonresonant excitation in this work, whose technical details are given in Appendix~B, above-bandgap light creates free carriers which pair up to form free excitons, which in turn are captured by T centers, as well as other defects in the sample. 
Once bound to a T center, the bound exciton recombines and with some probability (given by the quantum radiative efficiency) this process generates a photon. 
Of these radiative cases, with some probability (given by the Debye-Waller factor) this process generates a photon in the ZPL without any accompanying vibrational excitations (phonons or LVM), and the remainder of the time it produces vibrational excitations as well as light in the phonon sideband. 

\begin{figure}
    \includegraphics[width=\linewidth]{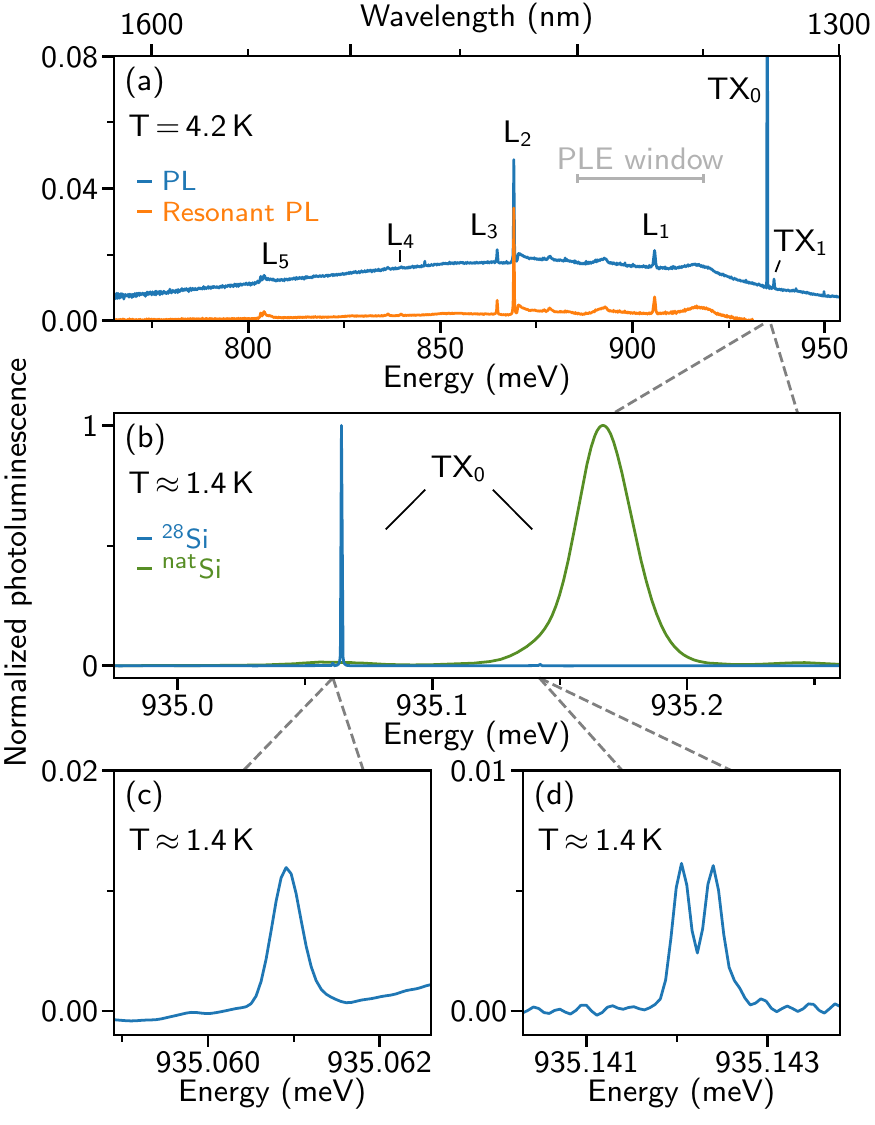}
    \caption{(a) PL spectra of the T center using above-gap and resonant excitation. The detection window used for PLE experiments in Sections \ref{sec:ZPL Temperature Dependence}, \ref{sec:Continuous-wave spin resonance} and \ref{sec:Pulsed spin resonance} is shown in grey. (b) High-resolution PL spectra of the T center TX$_0$ ZPL in $^\text{nat}$Si and in enriched $^{28}$Si. Weak features in the $^{28}$Si spectrum attributed to $^{13}$C isotope-shifted replicas for the T center's two inequivalent carbon atoms are visible at (c) 935.06093(7)\,meV, with an integrated area ratio of 1.1(1)\% relative to the TX$_0$ ZPL, and (d) a doublet at 935.14204(10)\,meV and 935.14240(6)\,meV, with a combined integrated ratio of 1.16(9)\% relative to the TX$_0$ ZPL. 
    \label{fig:photolum}
    }
    \end{figure}

In Fig.~\ref{fig:photolum} we report PL spectra of T centers in silicon. 
A low-resolution view of the T center ZPL and phonon sideband is shown in Fig.~\ref{fig:photolum}(a), which includes sharp features like the LVM replica labeled L$_2$ following \cite{Safonov1996}. 
In Fig.~\ref{fig:photolum}(b--d) we show high resolution spectra around the TX$_0$ ZPL.
The $^{\textrm{nat}}$Si spectrum shown in Fig.~\ref{fig:photolum}(b) is from the brightest T center sample in our possession, which was electron irradiated and heat treated in 1995 and stored at room temperature since then. Additional sample details are presented in Appendix~A.

In Fig.~\ref{fig:photolum}(a) the phonon sideband relative to the normalized ZPL is shown. 
When examining radiation-damaged silicon material in PL with non-selective above-gap excitation one unavoidably generates light from a variety of defects. 
These other defects have contributed in varying degrees to the backgrounds of all PL spectra of T centers reported to date, which make the extraction of the true Debye-Waller factor unreliable. 
The Debye-Waller factor can be measured accurately by normalizing above-bandgap PL spectra to resonantly excited PL spectra as described in Ref.~\cite{SisterPRL}.

In Fig.~\ref{fig:photolum}(b) we report the first high resolution optical spectroscopy of the TX$_0$ ZPL of T centers in $^{28}$Si. 
As has been observed with group-V donors \cite{Karaiskaj2001, Salvail2015}, group-VI double donors \cite{Morse2017, Deabreu2019} and other radiation damage centers \cite{Chartrand2018}, the removal of inhomogeneous local isotopic variations in the silicon lattice has a significant effect on the observed linewidth of the T center ZPL in isotopically enriched $^{28}$Si. 
In high-purity $^\text{nat}$Si we observe a linewidth of 26.9(8)\,$\upmu$eV, whereas in $^{28}$Si the linewidth is below the 0.25\,$\upmu$eV lower resolution limit of our FTIR spectrometer. 
A higher-resolution temperature-dependent linewidth study using PLE techniques presented in Section~\ref{sec:ZPL Temperature Dependence} reveals $^{28}$Si ensemble ZPL linewidths to be as low as 0.14(1)\,$\upmu$eV (33(2)\,MHz) -- nearly a 200-fold reduction in linewidth. 
In addition to a narrowing of the TX$_0$ line, we also observe a peak wavelength shift from 935.167(2)\,meV in $^\text{nat}$Si to 935.0643(1)\,meV in $^{28}$Si at 1.4\,K.
Similar spectral shifts, due to the dependence of the band-gap on average isotopic mass, electron-phonon coupling, and the defect binding energy, have been widely documented in silicon \cite{Karaiskaj2001}, including in work studying T centers in $^{30}$Si \cite{Hayama2004}. 

A closer look at the 1.4\,K high resolution $^{28}$Si PL spectra in Fig.~\ref{fig:photolum}(b) reveals the presence of a negatively-shifted satellite at 935.06093(7)\,meV (Fig.~\ref{fig:photolum}(c)) and a positively-shifted doublet at 935.14204(10)\,meV and 935.14240(6)\,meV (Fig.~\ref{fig:photolum}(d)) with integrated area ratios relative to the main ZPL of 1.1(1)\% and 1.16(9)\%, which are consistent with the 1.1\% natural isotopic abundance of $^{13}$C and the fact that the T center contains two inequivalent carbon atoms. 
Both features provide new understanding afforded by the isotopically purified $^{28}$Si host material. 
The negatively shifted $^{13}$C peak is fully obscured in $^\text{nat}$Si. The positively shifted feature, although characterised at low resolution in earlier isotopic composition studies \cite{Irion1985, Safonov1996}, reveals previously hidden structure: an 86(20)\,MHz splitting that is unobservable in $^\text{nat}$Si and is possibly connected to the $^{13}$C nuclear spin. 

\section{ZPL Temperature Dependence}
\label{sec:ZPL Temperature Dependence}

\begin{figure}
    \includegraphics[width=\linewidth]{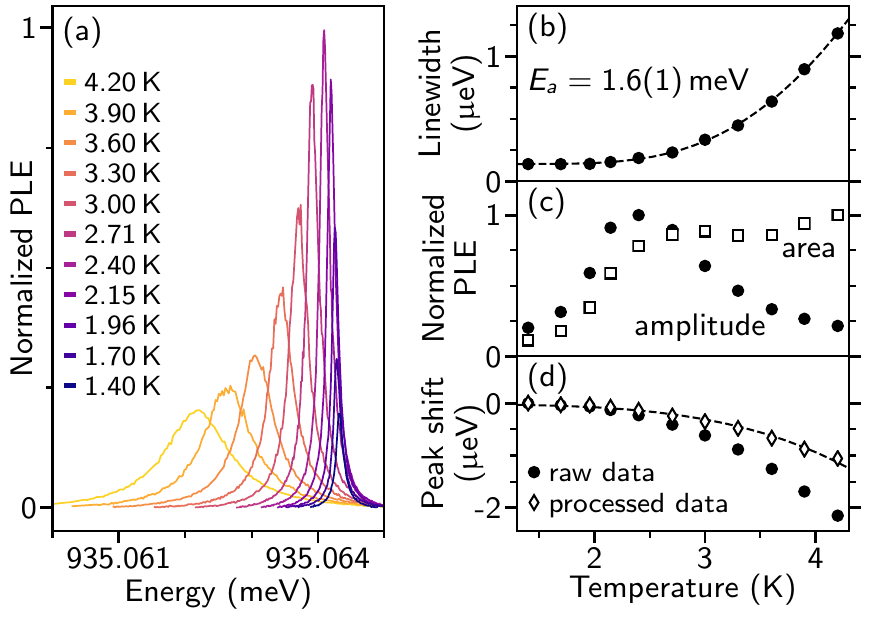}
    \caption{(a) Photoluminescence excitation (PLE) spectra of the TX$_0$ ZPL transition as a function of temperature, normalized to the 2.4\,K spectrum. The decrease in PLE signal at low temperatures indicates a shift from homogeneous to inhomogeneous broadening, as discussed in the text. (b--d) Lorentzian fit parameters for the spectra in panel (a). Error bars are smaller than the size of the data points. (b) The ZPL linewidth increases as a function of temperature following a thermally-induced transition model with an activation energy of 1.6(1)\,meV (details in text). (c) Normalized ZPL amplitude (black circles) and integrated area (white squares) as a function of temperature. (d) Peak position shift, where the black circles show the raw data and white diamonds show the shift when corrected for hydrostatic pressure effects.}
    \label{fig:tempdep}
    \end{figure}

PLE experiments with tunable single-frequency lasers, as described in Appendix~B, can measure linewidths below the resolution limits of high resolution spectrometers used for PL spectroscopy. 
Here we scan a laser over the TX$_0$ ZPL at different sample temperatures and detect sideband photons with wavelengths within the PLE detection window shown in Fig.~\ref{fig:photolum}(a).
The raw data as well as the fit results are shown in Fig.~\ref{fig:tempdep}. 

Lorentzian lineshapes provide an excellent fit to the raw data across the entire temperature range studied, giving linewidths of 1.18(3)\,$\upmu$eV at 4.2\,K and 0.14(1)\,$\upmu$eV at 1.4\,K.  
Fit parameters for all temperatures are plotted in Fig.~\ref{fig:tempdep}(b--d). The linewidth versus temperature data was fitted to a thermally-induced transition model \cite{Chartrand2018}: 
\begin{equation}
\label{eq:tempdep}
\Gamma(T) = P_0 + \frac{P_T}{\exp\left(E_a/k_BT\right) - 1}
\end{equation}

\noindent yielding $P_0$\,=\,0.137(8)\,$\upmu$eV, $P_T$\,=\,78(12)\,$\upmu$eV and an activation energy $E_a$\,=\,1.6(1)\,meV.
This is in reasonable agreement with the 1.76(1)\,meV TX$_0$ to TX$_1$ state splitting measured in PL at 4.2\,K. This demonstrates that the thermal broadening of the TX$_0$ line is due to thermally activated transitions between TX$_0$ and TX$_1$.

\begin{figure*}[t]
  \begin{minipage}{0.67\textwidth}
     \includegraphics[width=\textwidth]{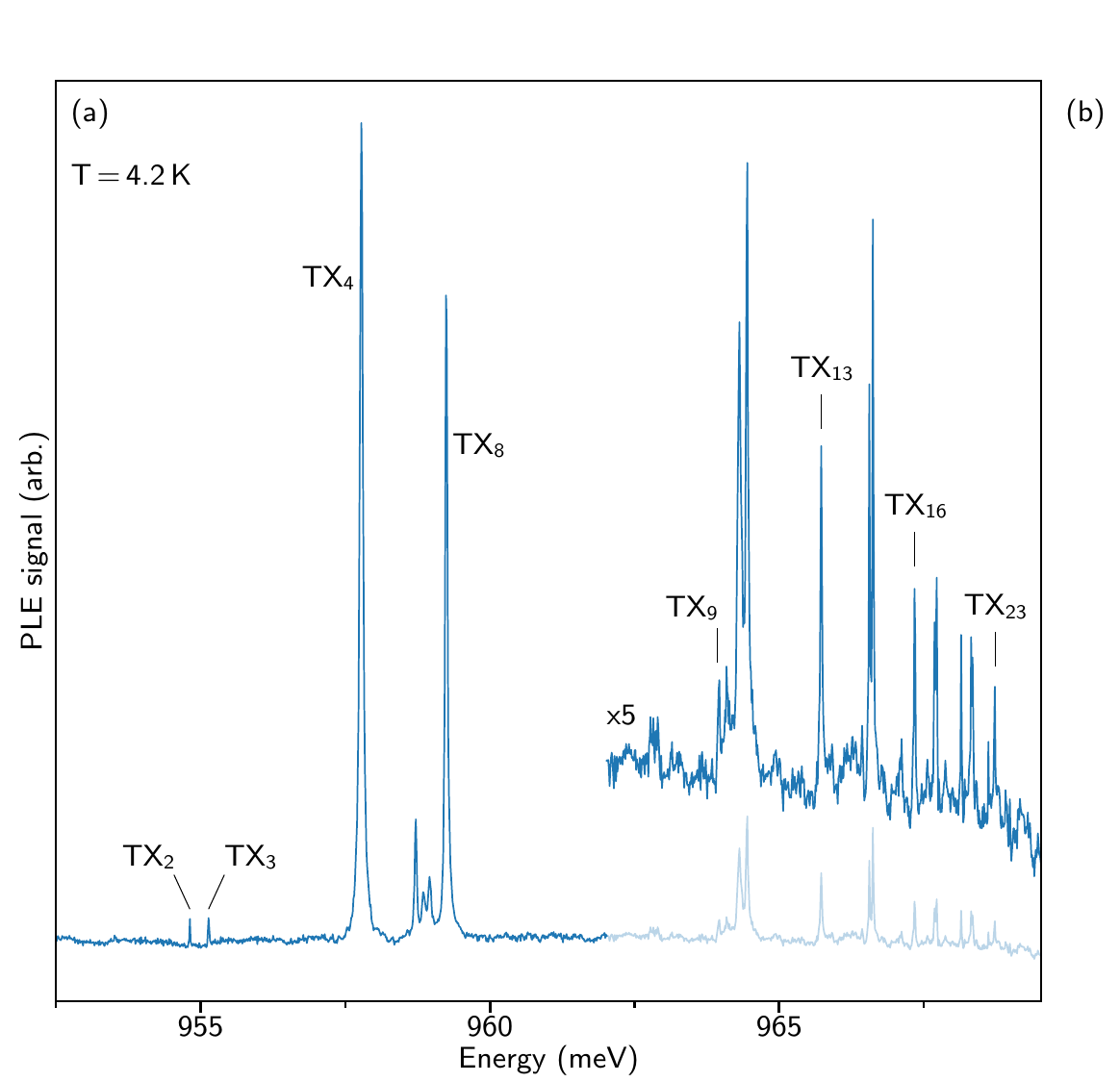}
  \end{minipage}%
  \begin{minipage}{0.33\textwidth}  
     \begin{tabularx}{\textwidth}{l@{\hspace{.5em}} r@{\hspace{1.3em}} r@{\hspace{1.3em}} r}
  \hline \hline\noalign{\smallskip}
    \textsf{Line}   & \textsf{Energy (meV)}  & \textsf{Amplitude}  & \textsf{State} \\
  \hline \noalign{\smallskip}
\textsf{TX$_\textsf{0}$   } & \textsf{935.0622} & \textsf{(PL 4.2~K)} & $\mathsf{1 \Gamma_8^+}$ \\
\textsf{TX$_\textsf{1}$   } & \textsf{936.82} & \textsf{(PL 4.2~K)} & $\mathsf{1 \Gamma_8^+}$\\
\textsf{TX$_\textsf{2}$   } & \textsf{954.81} & \textsf{0.026} &  \\
\textsf{TX$_\textsf{3}$   } & \textsf{955.14} & \textsf{0.027} & \\
\textsf{TX$_\textsf{4}$   } & \textsf{957.78} & \textsf{1.000} & $\mathsf{2 \Gamma_8^+}$\\
\textsf{TX$_\textsf{5}$   } & \textsf{958.72} & \textsf{0.147} & \\
\textsf{TX$_\textsf{6}$   } & \textsf{958.85} & \textsf{0.058} & \\
\textsf{TX$_\textsf{7}$   } & \textsf{958.96} & \textsf{0.077} & \\
\textsf{TX$_\textsf{8}$   } & \textsf{959.25} & \textsf{0.790} & $\mathsf{2 \Gamma_8^+}$\\
\textsf{TX$_\textsf{9}$ } & \textsf{963.96} & \textsf{0.025} & \\
\textsf{TX$_\textsf{10}$} & \textsf{964.09} & \textsf{0.028} & \\
\textsf{TX$_\textsf{11}$} & \textsf{964.31} & \textsf{0.112} &  $\mathsf{3 \Gamma_8^+}$\\
\textsf{TX$_\textsf{12}$} & \textsf{964.45} & \textsf{0.151} &  $\mathsf{3 \Gamma_8^+}$\\
\textsf{TX$_\textsf{13}$} & \textsf{965.73} & \textsf{0.082} &  $\mathsf{1 \Gamma_6^+}$\\
\textsf{TX$_\textsf{14}$} & \textsf{966.56} & \textsf{0.097} &  $\mathsf{4 \Gamma_8^+}$\\
\textsf{TX$_\textsf{15}$} & \textsf{966.62} & \textsf{0.137} &  $\mathsf{4 \Gamma_8^+}$\\
\textsf{TX$_\textsf{16}$} & \textsf{967.34} & \textsf{0.047} &  $\mathsf{2 \Gamma_6^+}$\\
\textsf{TX$_\textsf{17}$} & \textsf{967.68} & \textsf{0.039} &  $\mathsf{5 \Gamma_8^+}$\\
\textsf{TX$_\textsf{18}$} & \textsf{967.72} & \textsf{0.050} &  $\mathsf{5 \Gamma_8^+}$\\
\textsf{TX$_\textsf{19}$} & \textsf{968.14} & \textsf{0.036} &  $\mathsf{3 \Gamma_6^+}$\\
\textsf{TX$_\textsf{20}$} & \textsf{968.32} & \textsf{0.035} &  $\mathsf{6 \Gamma_8^+}$\\
\textsf{TX$_\textsf{21}$} & \textsf{968.34} & \textsf{0.030} &  $\mathsf{6 \Gamma_8^+}$\\
\textsf{TX$_\textsf{22}$} & \textsf{968.61} & \textsf{0.010} &  $\mathsf{4 \Gamma_6^+}$\\
\textsf{TX$_\textsf{23}$} & \textsf{968.73} & \textsf{0.023} &  $\mathsf{7 \Gamma_8^+}$\\
  \hline \hline  
    \end{tabularx}
    \end{minipage} 
\caption{(a) PLE spectrum revealing higher excited states of the T center bound exciton level TX. (b) Transition energies of all the known TX states, where TX$_0$ and TX$_1$ are determined using PL, and TX$_{>1}$ are determined from the PLE data shown in (a). The PLE peak amplitudes are shown relative to that of the TX$_\textsf{4}$ transition. The N$\Gamma_8^+$ state labels  for N$<$7 refer to the fourfold degenerate shallow acceptor states from which the two doubly degenerate TX states originate. This splitting can no longer be resolved for 7$\Gamma_8^+$ (TX$_\textsf{23}$). 
         \label{fig:excited} }  
\end{figure*}

Evidence for the transition from a thermally-broadened homogeneous line to an inhomogeneous line includes the PLE signal amplitude and integrated area decrease at temperatures below 2.4\,K as seen in Fig.~\ref{fig:tempdep}(c). 
A similar signal decrease is not observed in PL. 
In an inhomogeneously broadened line, local degrees of freedom such as spin states can shift the position of a given center's ZPL. 
If resonant optical driving changes these degrees of freedom, for example by flipping a spin state, that center's cycling transition frequency may no longer be in resonance with the optical driving frequency and the number of emitted photons will drop accordingly. 
This effect increases as the ratio of homogeneous to inhomogeneous linewidth decreases with lower temperatures. 
The lower bound on the homogeneous linewidth, from the temperature-independent lifetime data given in the sister publication to this work \cite{SisterPRL}, is $1/(2\pi\times 940\text{\,ns})$\,=\,0.169(2)\,MHz (or 0.700(8)\,neV).
The observation that an inhomogeneously broadened line is Lorentzian is not uncommon and could be due to strain from residual impurities, random electric fields from ionized impurities, or crystal damage \cite{Stoneham2001}. 
Further data in support of this interpretation are given in Section \ref{sec:Continuous-wave spin resonance}.

The peak position also changes with temperature as seen in Fig.~\ref{fig:tempdep}(d). 
The peak position shift includes both hydrostatic pressure and temperature shift effects, since the temperature is controlled by reducing the gas pressure over the liquid helium. As is the case with many silicon defects, the sample's ambient pressure should have a linear effect on the peak position because of the bandgap energy shift \cite{Cardona2004}. 
To measure the constant governing this linear shift, we compared two spectra -- one in liquid $^4$He, the other in gas $^4$He -- and made sure that they were at the same temperature by comparing their low-power (unsaturated) linewidths, which assumes that pressure changes have a negligible effect on the ZPL linewidth. 
Using this technique, the resulting constant is 1.47(34)$\times 10^{-9}$\,eV/Torr, which we subtracted from the total peak position shift. 
The remaining peak shift is well fitted by a $-$3.58\,$\upmu$eV/T$^4$ temperature dependence, which is proportional to the known T$^4$ behavior of the bandgap energy in this temperature range \cite{Cardona2004}.

\section{Bound Exciton Excited States}
\label{sec:Bound Exciton Excited States}

In the neutral ground state the T center contains an unpaired electron. 
During the lifetime of the bound exciton the TX center can be thought of as a pseudo-acceptor. 
Characteristics of this pseudo-acceptor include the existence of effective-mass-like excited states for the hole not unlike those of group III acceptors such as boron. 
The excited states of acceptors can be observed using a range of techniques \cite{Morse2016}. In the case of the T center's BE, a number of the excited states can be mapped out directly with PLE techniques. 

In the data presented in Fig.~\ref{fig:excited}, a tunable single-frequency laser was scanned above the energy of the TX$_{0,1}$ states, which resonantly populated bound excitons into various excited states of the center. 
We measured this excitation by TX$_0$ ZPL luminescence, resulting from phonon cascade from the excited state down to the TX$_0$ state, using a spectrometer and a single-photon detector as described in Appendix~B. 
To our knowledge this is the first reporting of the higher bound exciton excited states of the T center defect in silicon.

\begin{figure*}[t]
    \includegraphics[width=\textwidth]{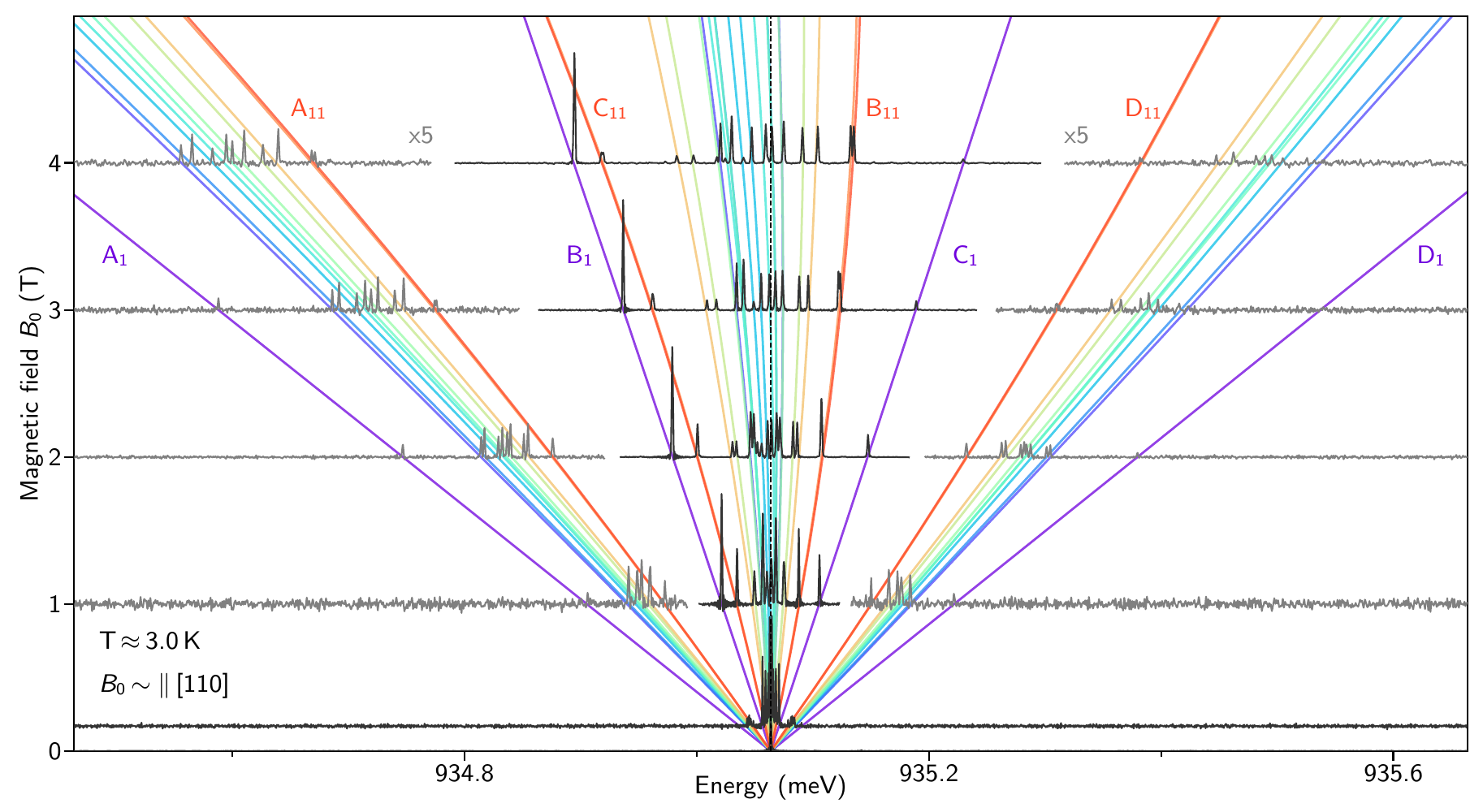}
    \caption{Photoluminescence of the T center TX$_0$ ZPL under an applied magnetic field with B$_0$ approximately parallel to the [110] crystal axis. The four optical transitions available to each identifiable orientational subset are grouped by color. Orientational subset 1 (purple) has optical transitions labeled A$_1$, B$_1$, C$_1$, D$_1$ in accordance with Fig.~\ref{fig:diagram}(c). Also labeled are the transitions of the subset having the smallest hole g factor, A$_{11}$, B$_{11}$, C$_{11}$ and D$_{11}$. The colored lines are fits to Eq.~\ref{eq:zeeman}, with resulting fit parameters given in Table~\ref{tab:g-factors}. A multiplier of $\times$5 has been applied to the wings of the spectra to enhance the visibility of lower amplitude lines.}
    \label{fig:magneto-pl}
    \end{figure*}

The TX  state energy spacings agree well with the known even parity states of shallow acceptors such as boron \cite{Morse2016}, with the complication that the normally fourfold degenerate N$\Gamma_8^+$ are split into two doubly degenerate states by the reduced symmetry local field of the defect. 
This size of this splitting decreases rapidly with increasing N, and is no longer resolved for $7\Gamma_8^+$ (TX$_\textsf{23}$). 
The relatively strong transitions labelled N$\Gamma_6^+$ again agree well with the level spacing known for the boron shallow acceptor \cite{Morse2016}. 
Plotting the transition energy vs.~$1/\textrm{N}^2$ for the N$\Gamma_8^+$ transitions shows the expected almost linear dependance of binding energy vs.~the inverse of the principal quantum number squared. 
This extrapolates to an ionization limit for the hole in TX at \mbox{969.84(15)~meV}, or a hole binding energy of 34.78(15) meV. 
The weaker unidentified transitions seen in Fig.~\ref{fig:excited}(a) could be weakly allowed transitions to odd-parity excited states, again possibly split by the defect field, although it cannot be ruled out that the lowest energy excited states TX$_\textsf{2}$ or TX$_\textsf{3}$ involve the shallow acceptor $1\Gamma_7^+$ state \cite{Morse2016}.

\section{Magneto-photoluminescence}
\label{sec:Magneto-photoluminescence}

The substantial reduction in optical inhomogeneous broadening afforded by the $^{28}$Si host material encouraged the re-examination of the PL of T centers under an applied magnetic field. 
As seen in Fig.~\ref{fig:magneto-pl}, the PL spectrum is substantially richer in $^{28}$Si than those of previous reported spectra in $^\text{nat}$Si \cite{Irion1985}, even in the moderate magnetic fields explored here. 

From the known C$_{\textrm{1h}}$ symmetry of the center, we expect twelve possible orientational subsets of T centers relative to an arbitrary B$_0$ field axis \cite{Kaplyanskii1967}. 
The unpaired \mbox{spin-1/2} electron was known to be highly isotropic in contrast with the highly anisotropic \mbox{spin-1/2} hole. 
Together this implies up to 48 observable optical transitions for a low-symmetry field axis, where each T center orientational subset gives rise to four allowed optical transitions, as schematically shown in Fig.~\ref{fig:diagram}(b). 
Further optical substructure, for example due to nuclear spin Zeeman and hyperfine effects, could not be resolved directly in these ensemble measurements.

Fig.~\ref{fig:magneto-pl} reveals eleven T center orientational subsets which are identified by color. 
Here we define `orientational subset 1': an orientational subset addressable with optical signals A$_1$, B$_1$, C$_1$, D$_1$, which underpins the spin resonance studies in Sections \ref{sec:Continuous-wave spin resonance} and \ref{sec:Pulsed spin resonance}. 
Furthermore, in Section \ref{sec:Continuous-wave spin resonance} we reveal through magnetic resonance techniques that the purple subset with assigned A$_1$, B$_1$, C$_1$, D$_1$ labels in Fig.~\ref{fig:magneto-pl}~addresses two near-degenerate orientational subsets, bringing the total observed number of orientational subsets in this work to twelve.
Note that subset 1 has the largest effective hole g factor of all the subsets. 
We also label the transitions of the subset having the smallest hole g factor as A$_{11}$, B$_{11}$, C$_{11}$ and D$_{11}$.  
Note that the ABCD labels indicate transitions connecting specific spin states, and not the energetic ordering of the transitions, so the energy of B vs. C is reversed for subsets 4 through 11. 

Each orientational subset's optical transition frequencies are well fit to a spin Hamiltonian model with Zeeman terms for each of the unpaired ground state electron spin and excited state hole spin as well as a diamagnetic term. Specifically, we fit using the following set of four equations: 
\begin{equation}
\label{eq:zeeman}
\{\text{A,\,B,\,C,\,D}\} = \pm \frac{g_\textrm{E} \mu_\textrm{B} | {\bf B_0} | }{2e} \pm \frac{g_\textrm{H} \mu_\textrm{B} | {\bf B_0} |}{2e}  + \chi | {\bf B_0} |^2
\end{equation}

\noindent where \{A,\,B,\,C,\,D\} is the peak position shift from its zero field value, $g_\textrm{E}$ is the ground state electron g factor constant for a given orientational subset, $\mu_\textrm{B}$ is the Bohr magneton, ${\bf B_0}$ is the magnetic field vector, $e$ is the elementary charge, $g_\textrm{H}$ is the effective bound exciton hole g factor for a given orientational subset, and $\chi$ is the diamagnetic shift. 
All ground state electron g factors are equal within error, with an average of 2.005(8). Results of the fit for the effective hole g factors and diamagnetic shifts of all identifiable subsets are given in Table~\ref{tab:g-factors}.

\begin{table}
	\centering
  \begin{tabularx}{\linewidth}{r@{\hspace{0.5em}}l@{\hspace{1em}} r@{\hspace{1.3em}} r}
  \hline \hline\noalign{\smallskip}
      &\multicolumn{2}{r@{\hspace{1.3em}}}{Effective TX$_0$ hole spin g factor} & Diamagnetic shift \\
      && $g_\textrm{H}$ & $\chi$ ($\unslant\mu$eV/T$^2$) \\
  \hline \noalign{\smallskip}
1  &\color{rainbow0}$\blacksquare$&3.457(7) &$-$0.11(2)\\ 
2  &\color{rainbow1}$\blacksquare$    &2.233(9)     &$-$1.05(6)    \\ 
3  &\color{rainbow2}$\blacksquare$    &2.165(14)    &$-$1.03(6)    \\ 
4  &\color{rainbow3}$\blacksquare$    &1.970(12)    &$-$1.26(4)    \\ 
5  &\color{rainbow4}$\blacksquare$    &1.871(22)    &$-$1.36(9)    \\ 
6  &\color{rainbow5}$\blacksquare$    &1.851(14)    &$-$1.03(6)    \\ 
7  &\color{rainbow6}$\blacksquare$    &1.770(8)     &$-$1.10(7)    \\ 
8  &\color{rainbow7}$\blacksquare$    &1.596(6)     &$-$1.23(5)    \\ 
9  &\color{rainbow8}$\blacksquare$    &1.497(11)    &$-$1.26(9)    \\ 
10 &\color{rainbow9}$\blacksquare$    &1.082(7)     &$-$2.35(3)    \\ 
11 &\color{rainbow10}$\blacksquare$   &1.069(7)     &$-$2.35(3)    \\ 
  \hline \hline
  \end{tabularx}
	\caption{Effective TX$_0$ hole spin g factors and diamagnetic shift constants for the different T center orientational subsets observed with this magnetic field axis (approximately parallel to the [110] crystal axis), whose colors match the subsets labeled in Fig.~\ref{fig:magneto-pl}.
    \label{tab:g-factors}
    }
\end{table}

The relative amplitudes of the four lines associated with a given orientational subset measured using PL are affected by: i) spin selection rules, and ii) the hole spin thermal populations, where we observe that the hole spin relaxation time is comparable to the TX BE lifetime of 940~ns \cite{SisterPRL}.
As discussed in the next section, the spectra obtained by PLE in a magnetic field are quite different from those obtained using PL.

\FloatBarrier
\section{Continuous-wave spin resonance}
\label{sec:Continuous-wave spin resonance}

    \begin{figure}
    \includegraphics[width=\linewidth]{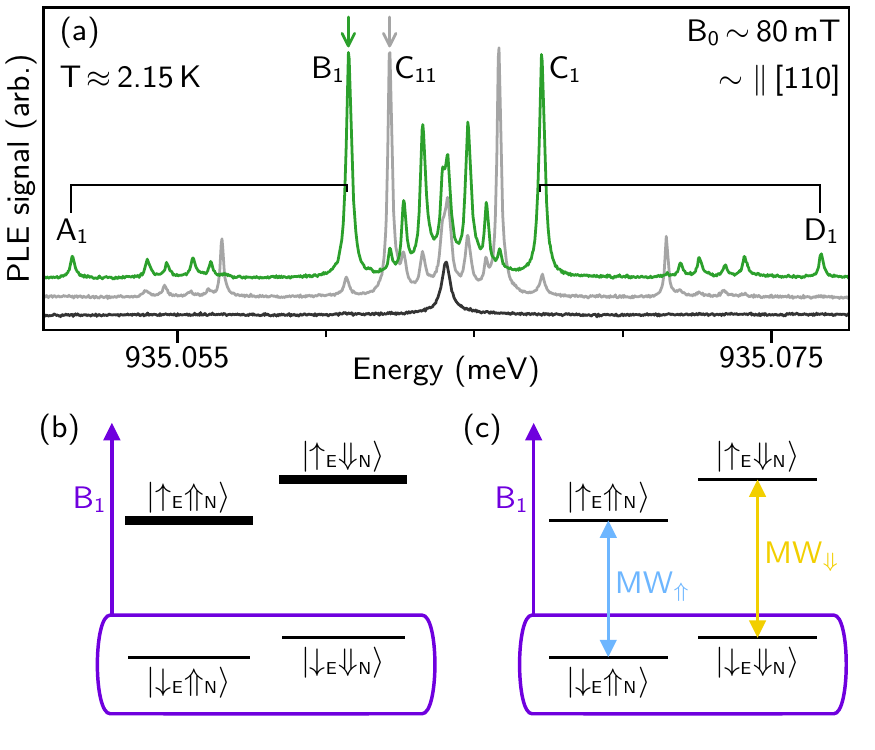}
    \caption{(a) PLE spectrum of the TX$_0$ ZPL under a static magnetic field of $\sim$80\,mT oriented approximately parallel to the [110] crystal axis, with and without electron spin depolarization. Line labels A, B, C, D match those of Fig.~\ref{fig:diagram}(c) and \ref{fig:magneto-pl}. Brackets display the g$_\textrm{E}$\,=\,2.005(8) electron splitting at this field. The black trace (bottom) was collected without any MW excitation. The green trace (top) was obtained when pumping MW$_\Uparrow$\,=\,2.2504 GHz and MW$_\Downarrow$\,=\,2.2533 GHz, which maximizes the signal from subset 1. The grey trace (middle) includes a single central MW signal which drives both MW$_\Uparrow$ and MW$_\Downarrow$ from subset 11. (b and c) Ground state energy levels created by the hyperfine interaction between the electron spin and the nuclear spin, as per Eq. \ref{eq:ham}. The schematic level diagrams picture electron spin hyperpolarization in (b) and electron spin mixing in (c). Shelving is indicated by a thicker energy level.}
    \label{fig:magneto-ple}
    \end{figure}

The many optical transitions seen in magneto-photoluminescence are not immediately observable under the resonant optical driving of a PLE measurement. 
As shown in Fig.~\ref{fig:magneto-ple}(a, bottom trace), without the application of any external signals, only a small central peak is visible under an applied magnetic field of 80\,mT when using the optically detected magnetic resonance apparatus described in Appendix~B. 
The absence of most optical transitions in PLE is an expected consequence of ground state electron spin hyperpolarization resulting from spin-selective resonant excitation, as shown schematically in Fig.~\ref{fig:magneto-ple}(b). 
Under such a model the central peak remains visible because for a few orientational subsets the ground and exciton spin states share approximately the same effective g factor value, which is to say that their B and C transitions almost overlap, and so both electron ground spin states can be continually optically excited at that central frequency and electron spin hyperpolarization does not occur.

To regain some of the structure observed in magneto-PL, resonant continuous-wave (CW) electron paramagnetic resonance (EPR) in the form of microwave (MW) irradiation can be applied to continually depolarize or mix the electron spin, as shown schematically in Fig.~\ref{fig:magneto-ple}(c). 
Spin mixing depopulates the spin shelving state(s) which are not being optically pumped and allows the optical excitation cycle to resume. 

Overall the spin Hamiltonian for the T ground state $\mathcal{H_T}$ with two $^{12}$C constituents is given by
\begin{equation}
\mathcal{H_T} = \mu_\textrm{B} {\bf B_0 g_E  S} + \mu_\textrm{N} g_\textrm{N} {\bf B_0 I} + h {\bf S  A I} \label{eq:ham}
\end{equation}
where $\mu_\textrm{B}$ is the Bohr magneton, ${\bf B_0}$ is the magnetic field vector, ${\bf g_E}$ is the electron spin g factor tensor which is approximately isotropic with ${g_\textrm{E} = 2.005(8)}$, $\bf{S}$ is the electron spin vector, $\mu_\textrm{N}$ is the nuclear spin magneton, ${g_\textrm{N}}$ is the hydrogen nuclear spin g factor, $\bf{I}$ is the hydrogen nuclear spin vector, $h$ is the Planck constant, and $\bf{A}$ is the hyperfine tensor. 
We find $\bf{A}$ is approximated well by a constant $A_{\textrm{eff}}$ for a particular orientational subset and field vector to obtain the results in this work. 
This spin Hamiltonian gives rise to the level structure shown in Fig.~\ref{fig:magneto-ple}(b and c) under our experimental conditions, where we label the two conditional MW frequencies MW$_\Uparrow$ and MW$_\Downarrow$ according to their respective nuclear spin state.

Even though we will find that these orientational subsets have small effective hyperfine splittings due to the nuclear spin of the hydrogen, a relatively strong MW field resonant with the g$_\textrm{E}$\,=\,2.005(8) transition frequency at $\sim$80\,mT is sufficient to simultaneously pump both MW$_\Uparrow$ and MW$_\Downarrow$ resonances thanks to power broadening, partially depolarizing the ground state electron spins and strengthening the central PLE component, as well as allowing many of the other PLE components to be observed, as shown in the middle trace (grey data) of Fig.~\ref{fig:magneto-ple}(a).
From this observation we can definitively conclude that the isotropic unpaired electron spin occurs in the unexcited T center, and not in the TX BE state.

    \begin{figure}
    \includegraphics[width=\linewidth]{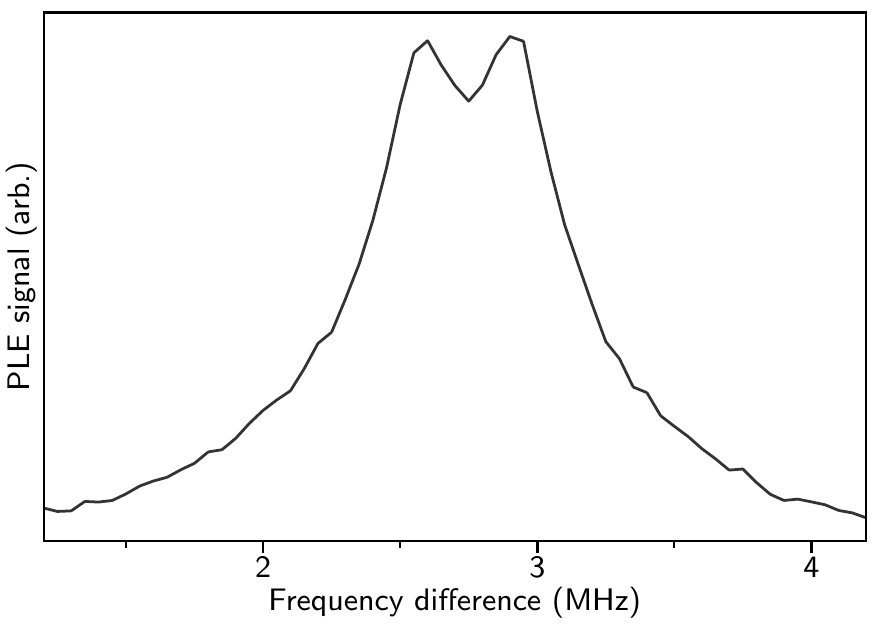}
    \caption{ODMR response at $\sim$80~mT when pumping optical transition B$_1$ and symmetrically sweeping two microwave frequencies outwards from the central MW frequency under these conditions, $g_\textrm{E} \mu_\textrm{B} |B_0|$ = 2.255\,GHz, plotted as a function of the difference between the two swept frequencies. The observed doublet depolarization response at difference frequencies of 2.57(1) MHz and 2.93(1) MHz reveals two distinct hyperfine splittings from two slightly different orientational subsets of subset 1, which are not optically resolvable.}
    \label{fig:hyperfine}
    \end{figure}

We find that for specific orientational subsets, specific pairs of microwave frequencies centered on the   g$_\textrm{E}$\,=\,2.005(8) value mix the ground state spins most effectively, maximizing the PLE signal from that orientational subset.
For example, the top trace in Fig.~\ref{fig:magneto-ple}(a), where the subset 1 signal is maximized with MW$_\Uparrow$\,=\,2.2504 GHz and MW$_\Downarrow$\,=\,2.2533 GHz, reveals an effective hyperfine splitting $|A_{\textrm{eff}}|$= 2.9 MHz for subset 1 (we will later show that closer study of subset 1 actually reveals two different values of $|A_{\textrm{eff}}|$).  
The differences between the top and middle trace of Fig.~\ref{fig:magneto-ple}(a) demonstrate that the hydrogen spin hyperfine interaction of the T center is anisotropic.

The specific MW frequency combination used to generate the green data in Fig.~\ref{fig:magneto-ple}(a) was chosen as a result of magnetic resonance spectroscopy experiments. This optimization process is described next.
 
For orientational subset $i$, the tunable laser was first set to a B$_i$ or C$_i$ transition energy as determined in Section \ref{sec:Magneto-photoluminescence}. 
For example, for orientational subset 1, this energy is indicated by the letter B$_1$ (green arrow) in Fig.~\ref{fig:magneto-ple}(a). 
In this case both the B$_1$ and D$_1$ optical transitions promote the T $\ket{\downarrow_\textrm{E}}$ electron states to TX$_0$, consequently hyperpolarizing the electron into the $\ket{\uparrow_\textrm{E}}$ electron state; the A$_1$ and C$_1$ optical transitions generate the opposite electron spin hyperpolarization. 
Two tunable MW sources were initially each set to the central MW frequency given by the electron spin g factor value. 
Following this, a symmetric MW frequency sweep was applied to reveal the pair of MW frequencies, split by the hydrogen hyperfine interaction, able to most effectively depolarize that particular orientational subset and generate an optically detected magnetic resonance (ODMR) signal. 

For a single orientational subset and accurately centered initial MW frequencies the result of this sweep should display a single peak at the effective hyperfine value associated with that orientational subset at the chosen magnetic field direction. 
The result of this scan applied to orientational subset 11, with the laser set to the C$_{11}$ optical transition (grey arrow in Fig.~\ref{fig:magneto-ple}(a)) reveals that the subset 11 PLE signal is maximized for a MW frequency difference of $|A_{\textrm{eff}}|$=0.40(3) MHz.
The result of this scan applied to orientational subset 1 is shown in Fig.~\ref{fig:hyperfine}.
The doublet observed in Fig.~\ref{fig:hyperfine} reveals that orientational subset 1 actually comprises a pair of orientational subsets hereafter referred to as subset 1 and subset 1$^\prime$, thus bringing the total number of orientational subsets to twelve.
A two-Lorentzian fit reveals effective hyperfine magnitudes $|A_{\textrm{eff}}|$ of 2.93(1)\,MHz and 2.57(1)\,MHz for subset 1 and subset 1$^\prime$ respectively. 
This has implications for the magnetic resonance experiments discussed later in this section. 
For the purpose of generating the green data in Fig.~\ref{fig:magneto-ple}(a), a single $|A_{\textrm{eff}}|$ near the average of these two values was used.

    \begin{figure}
    \includegraphics[width=\linewidth]{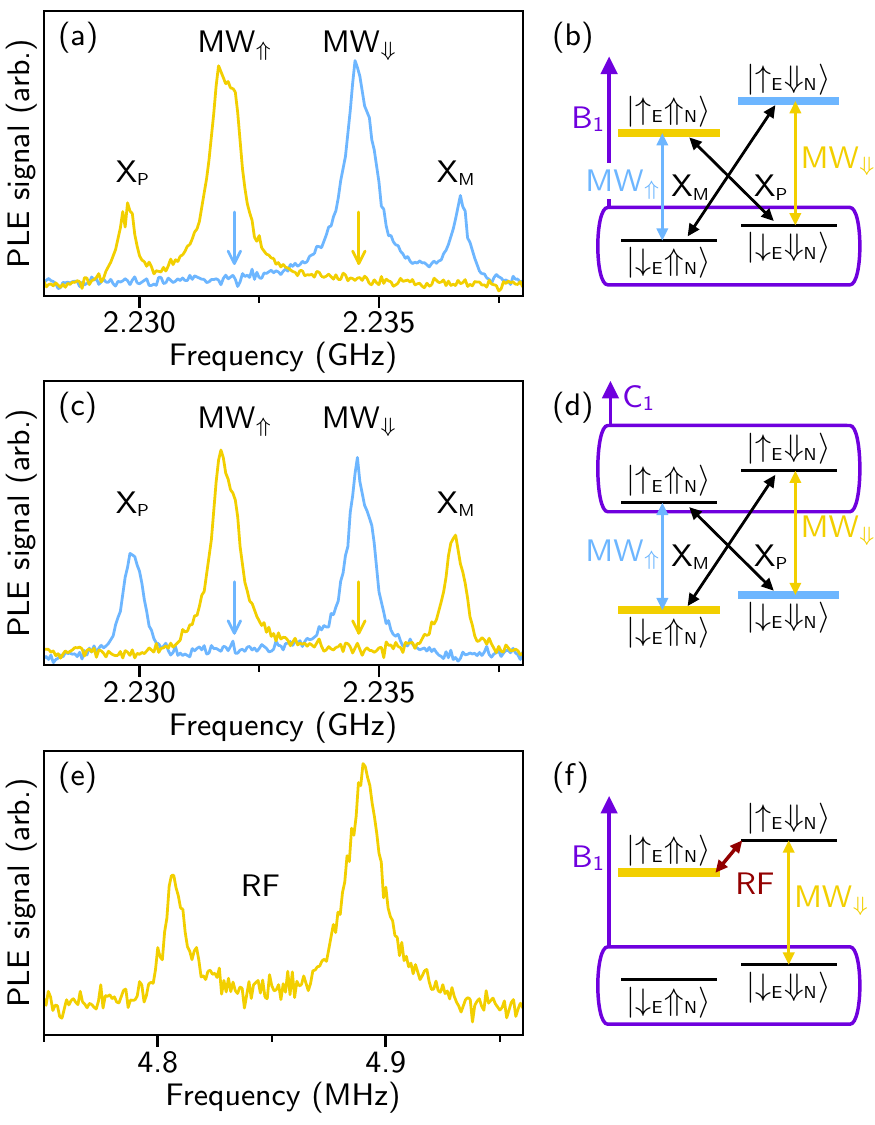}
    \caption{ODMR peaks reflecting a relaxation of the hyperpolarization generated by pumping a fixed optical and microwave transition are observed as a second MW or RF frequency is scanned across specific resonances. (a) ODMR intensity as a second MW source is scanned while pumping optical transition B$_1$ as well as either MW$_\Uparrow =~$2.23198 MHz (blue data and arrow) or MW$_\Downarrow =~$2.23458 MHz (yellow data and arrow). (b) Level scheme for (a) showing the shelving level for both fixed frequencies in the absence of depolarization as a thick appropriately colored line. (c) and (d) are as for (a) and (b) but while pumping optical transition C$_1$. (e) ODMR response while scanning an RF pump frequency across the region corresponding to the RF transition shown in (f) while also pumping optical transition B$_1$ as well as MW$_\Downarrow$.
}    \label{fig:eprnmr}
    \end{figure}

Starting from the average $|A_{\textrm{eff}}|$ for subsets 1 and 1$^\prime$ measured above, it was possible to refine the MW frequency estimates by fixing one MW frequency and sweeping the other. 
The results of this are seen in Fig.~\ref{fig:eprnmr}(a) and \ref{fig:eprnmr}(c). 
In both instances, if only one MW transition was being resonantly driven no substantial ODMR signal was observed, yet if both `allowed' MW transitions were being driven the ODMR signal was recovered. 
Equivalently put, both the electron and nuclear spins can be efficiently hyperpolarized using a single optical frequency and a selectively resonant MW frequency, similar to what has been reported for excitons bound to shallow donors \cite{Steger2011}. 
Interestingly, the ODMR signal could also be recovered by driving the so-called `forbidden' MW transitions, labelled $X_\textrm{M}$ (for mixed) and $X_\textrm{P}$ (for pure) in Fig.~\ref{fig:eprnmr}, which can be driven in systems with an anisotropic hyperfine interaction \cite{Schweiger2001}.

Notably, the hyperpolarization mechanism for T centers is not presently believed to include a substantial Auger bound-exciton recombination component as is known to be the case for the shallow donors. 
Efforts were made at zero magnetic field to observe a change in sample conductivity when the laser was applied on resonance with the ZPL, as was used with great success to indirectly measure the spin-dependent creation of excitons bound to shallow donors \cite{Steger2012}. 
For T centers these efforts did not reveal any evidence for Auger recombination whatsoever. 
Moreover we have not observed any optical bleaching effects by driving the TX$_0$ transition. 
Together this indicates that, with a high probability, an unpaired electron remains bound to the T center following bound exciton recombination.

Starting from a spin-polarized configuration obtained by applying both on-resonant optical B$_1$ and MW signals, we can perform nuclear magnetic resonance (NMR) by applying radio frequency (RF) signals resonant with the nuclear spin transition frequencies. 
The predicted hydrogen spin transition frequencies near 2.1\,MHz and 4.9\,MHz correspond to the cases where the hydrogen spin is coupled to the two different electron spin states. 
With MW$_\Downarrow$ continuously driven and the B$_1$ optical transition as chosen, which is pumping from the electron spin-down state, only the electron spin-up RF frequency should depolarize the otherwise hyperpolarized nuclear spin state and generate PLE signal. 
This configuration is shown in Fig.~\ref{fig:eprnmr}(f).
The sign of $A_{\textrm{eff}}$ determines whether the electron-spin-up RF frequency is near 2.1\,MHz or near 4.9\,MHz. 
In Fig.~\ref{fig:eprnmr}(e) we observe two RF transitions near 4.9\,MHz, corresponding to the two addressable orientational subsets 1 and 1$^\prime$ first identified in Fig.~\ref{fig:hyperfine}, which are more clearly distinguished here. 
From this we infer that $A_{\textrm{eff}}$ for these orientational subsets under these conditions are negative: $A_{\textrm{eff}}$\,=\,$-$2.93(1)\,MHz\ and $-$2.57(1)\,MHz. 
A negative $A_{\textrm{eff}}$ can be observed when the anisotropic dipolar hyperfine component reaches values larger than that of the isotropic contact hyperfine component of the overall hyperfine interaction \cite{Schweiger2001}. 

In this work the T centers under investigation are those with $^{12}$C constituents. Upcoming studies with $^{13}$C will offer four \mbox{spin-1/2} qubits per T center, with a correspondingly richer spin Hamiltonian for each orientational subset.

\section{Pulsed spin resonance}
\label{sec:Pulsed spin resonance}

    \begin{figure}[t]
    \includegraphics[width=\linewidth]{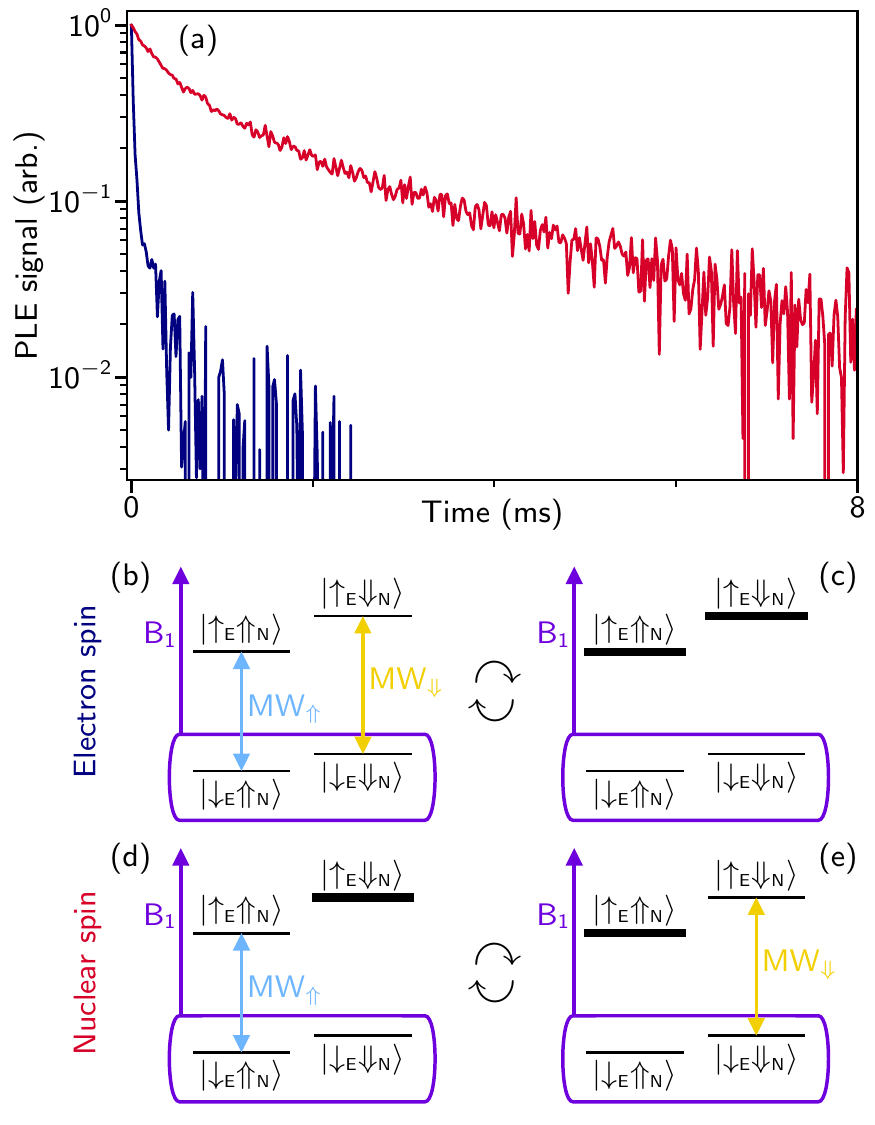}
    \caption{(a) Nuclear spin (red) and electron spin (dark blue) polarization transients. (b) and (c) Energy level diagrams illustrating the electron spin mixing (b) and polarization (c) transient schemes. (b) Populations are first prepared into a mixed population distribution by activating B$_1$ as well as MW$_\Uparrow$ and MW$_\Downarrow$. (c) To observe the electron spin polarization transient, the MW signals are halted. Shelving is indicated by a thicker energy level. (d) and (e): Energy level diagrams illustrating the nuclear spin polarization transient scheme. (d) Populations are first prepared into the $\ket{\uparrow_\textrm{E}\Downarrow_\textrm{N}}$ state by pumping B$_1$ and MW$_\Uparrow$. (e) To collect the transient, MW$_\Uparrow$ is turned off while MW$_\Downarrow$ is turned on, changing the shelving state from $\ket{\uparrow_\textrm{E}\Downarrow_\textrm{N}}$ to $\ket{\uparrow_\textrm{E}\Uparrow_\textrm{N}}$.
    \label{fig:transients}
    }
    \end{figure}

From Section \ref{sec:Continuous-wave spin resonance} we have determined spin Hamiltonian values for orientational subset 1 as well as techniques able to efficiently hyperpolarize both the electron and nuclear spin states. 
In this section, we introduce pulsed optical and magnetic resonance techniques which are able to manipulate and measure the T center spin populations.
These techniques are employed in the sister publication to this work \cite{SisterPRL} to extract spin T$_1$ times substantially longer than 16 seconds and Hahn-echo T$_2$ times exceeding a millisecond and a second for the electron and nuclear spins respectively.

All spin measurements are predicated upon having spin initialization and read out techniques. 
The electron-nuclear spin hyperpolarization process outlined in Section \ref{sec:Continuous-wave spin resonance}, whereby resonant optical and MW frequencies are applied to the entire ensemble, is used to initialize the spins. 
Next, we discuss pulsed optical techniques to read out the T center spin populations.

We begin with the electron spin polarization, which could be measured most easily by switching the optical pump signal back and forth between B$_1$, which pumps the system towards electron polarization $+1$ (all electron spins up), and C$_1$, which pumps the system towards electron polarization $-1$ (all electron spins down). In both cases the steady-state sideband luminescence approaches zero, since all systems are pumped into a shelving state which the optical pump does not access, but when switching from one pump to the other there will be an initial luminescence transient whose amplitude is proportional to the difference in the polarization state just before the pumping is switched and the steady state polarization which the pumping eventually produces. This scheme would require two single frequency lasers, so we used a different approach.

A mixed electron spin state (electron polarization $0$) can be produced by applying B$_1$, MW$_\Uparrow$, and MW$_\Downarrow$ simultaneously as shown in Fig.~\ref{fig:transients}(b). 
An electron spin polarization transient is produced by shutting off MW$_\Uparrow$ and MW$_\Downarrow$ and letting B$_1$ hyperpolarize the electron spin into $\ket{\uparrow_\textrm{E}}$, as shown in Fig.~\ref{fig:transients}(c). 
The transient luminescence shown in Fig.~\ref{fig:transients}(a) (blue data) is the signal generated when the pumping is switched from Fig.~\ref{fig:transients}(b) to Fig.~\ref{fig:transients}(c), or in other words the electron polarization is driven from $0$ to $+1$. 

In principle, the electron spin state can be deduced optically during this process by detecting the resulting sideband luminescence transient. 
However, this electron-spin-selective optical excitation cycle leads to very rapid electron spin hyperpolarization. 
The branching ratios from the BE states back down to the ground electron spin states are presumably relatively balanced, and the electron spin is hyperpolarized within a few optical cycles.  
As a result, very few sideband luminescence photons are collected during electron spin transient measurements.

Instead of measuring the electron spin using these transients, an indirect method measuring the nuclear spin state was employed. 
The nuclear spin luminescence polarization transient data is shown Fig.~\ref{fig:transients}(a) (red data), and the preparation and readout schemes are shown in Fig.~\ref{fig:transients}(d) and (e), respectively. 
The pumping in Fig.~\ref{fig:transients}(d) polarizes the system to the $\ket{\uparrow_\textrm{E}\Downarrow_\textrm{N}}$ state, while that in (e) polarizes it to the $\ket{\uparrow_\textrm{E}\Uparrow_\textrm{N}}$ state.  
The red luminescence transient data in Fig.~\ref{fig:transients}(a) was generated when the pumping switched from (d) to (e). 
A very similar transient is observed when the pumping is switched from (e) to (d).

What is noteworthy in Fig.~\ref{fig:transients}(a) is the much longer decay time, and thus integrated transient area, of the nuclear polarization transient (red) compared to the electron polarization transient (blue). 
This must result from the fact that the T centers can go through many optical absorption/emission cycles before the nuclear spin is flipped.  
The direct measurement of the electron polarization as in Fig. 9 has a much lower signal-to-noise ratio than the measurement of the nuclear polarization.  
In the following measurements the electron polarization is therefore measured indirectly, by mapping it onto the nuclear spin using an electron-spin-selective nuclear $\pi$ pulse.

\begin{figure}[t]
    \includegraphics[width=\linewidth]{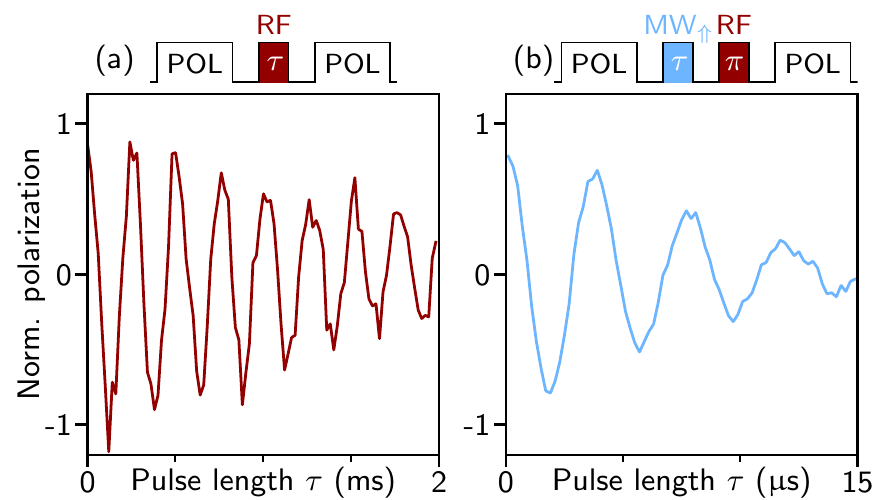}
    \caption{Rabi oscillations for (a) the nuclear spin and (b) the electron spin. The signal decay is limited by the EPR and NMR resonators' magnetic field inhomogeneities. The normalization is based off exponentially decaying cosine fits. The pulse sequence colors match the transition energies in the level diagrams shown in Figs.~\ref{fig:magneto-ple},~\ref{fig:eprnmr}, and ~\ref{fig:transients}. }
    \label{fig:rabi}
    \end{figure}

Furthermore, the alternating initialization and readout cycles used in Fig.~\ref{fig:transients} can be simplified by using a single polarization combination, in our case that shown in Fig. 9(e), which leaves the system in $\ket{\uparrow_\textrm{E}\Uparrow_\textrm{N}}$, to both detect any polarization transient and then re-initialize the system.  
This initialization / readout procedure is labelled `POL' in Fig.~\ref{fig:rabi}, and the pump laser is mechanically blocked in the interval between POL cycles.  
If the nuclear spins do not change between one cycle and the next, there will be no transient, and if the nuclear spin state polarization differs through decay or spin manipulation, a transient will be generated whose amplitude will be proportional to the fraction of centers having nuclear spin down.

In Fig.~\ref{fig:rabi}(a) we combine these methods to measure Rabi oscillations when driving the nuclear spins with variable length RF pulses at the electron spin up RF frequency.  
For zero RF pulse length there is no change in nuclear spin, so there is zero transient area, and the nuclear spin polarization is near $+1$, as it is at the end of the POL period.  
When the RF pulse length reaches $\pi$, the population is flipped to $\ket{\uparrow_\textrm{E}\Downarrow_\textrm{N}}$, and the nuclear polarization is near $-1$.

In Fig.~\ref{fig:rabi}(b) we measure the Rabi oscillations due to variable length MW pulses which drive the electron spin.  
As before, the system is initialized to $\ket{\uparrow_\textrm{E}\Uparrow_\textrm{N}}$ before the POL pumping is stopped and the spin pulse sequence is applied.  
The MW$_\Uparrow$ pulse applied to this state rotates the electron spins, and the RF $\pi$ pulse flips any remaining population within $\ket{\uparrow_\textrm{E}\Uparrow_\textrm{N}}$ to $\ket{\uparrow_\textrm{E}\Downarrow_\textrm{N}}$, generating a nuclear polarization transient when B$_1$ and MW$_\Downarrow$ are turned on again in the POL cycle.
For zero MW$_\Uparrow$ pulse length, the RF $\pi$ pulse flips all of the systems back to $\ket{\uparrow_\textrm{E}\Downarrow_\textrm{N}}$, and a maximum sideband luminescence transient area is observed, corresponding to electron spin polarization near $+1$.

In Fig.~\ref{fig:rabi} the normalization of the observed Rabi oscillations are determined from exponentially decaying cosine fits. The decay of the Rabi oscillations seen in Fig.~\ref{fig:rabi} are thought to be due to MW and RF field inhomogeneities, and the observed spin control fidelities are typical for ensemble measurements in home-built resonators.

\FloatBarrier
\section{Conclusion}
\label{sec:Conclusion}

We have reported upon the promising optical and spin properties of an ensemble of T centers in $^{28}$Si. We have performed above-bandgap PL, resonant PL, PLE, and ODMR upon this ensemble. 
As has been reported with other radiation damage centers' ZPL transitions in $^{28}$Si, we observe a significant (here nearly 200-fold) reduction in ensemble inhomogeneous linewidth, to as low as 33(2) MHz at $\sim$1.4~K. 
PLE traces at $\sim$1.4~K confirm that this ensemble linewidth is inhomogeneously broadened. 
We have observed two distinct carbon isotope shifts of the ZPL, which is consistent with existing atomic models of the T center. 
This opens up avenues for the future study of T centers with up to three \mbox{spin-1/2} nuclei. 

In a magnetic field, we observed twelve orientational subsets, consistent with the C$_\textrm{1h}$ symmetry of the center. 
The hyperpolarization dynamics observed in PLE in a magnetic field allowed for ODMR and the determination of MW and RF transition frequencies for a single orientational subset of T centers. 
From this we have proven that the unexcited ground state contains an unpaired electron spin, conclusively resolving this open question in the literature.
Furthermore, we have shown that there exists an anisotropic hyperfine interaction with the defect's hydrogen nuclear spin, suitable to support spin readout via the nuclear spin's slow hyperpolarization optical transient. 
Taken together these results pave the way for T centers to successfully hybridize silicon's two dominant quantum platforms, and provide a long-lived multi-qubit backbone for future telecom-wavelength integrated quantum photonic circuits. 

\section*{Acknowledgements}
This work was supported by the Natural Sciences and Engineering Research Council of Canada (NSERC), the Canada Research Chairs program (CRC), the Canada Foundation for Innovation (CFI), the B.C. Knowledge Development Fund (BCKDF), and the Canadian Institute for Advanced Research (CIFAR) Quantum Information Science program. 
The $^{28}$Si samples used in this study were prepared from the Avo28 crystal produced by the International Avogadro Coordination (IAC) Project (2004--2011) in cooperation among the BIPM, the INRIM (Italy), the IRMM (EU), the NMIA (Australia), the NMIJ (Japan), the NPL (UK), and the PTB (Germany).
We thank Valentin Karasyuk for fruitful discussions. 
We thank Alex English of Iotron Industries for assistance with electron irradiation.

\section*{APPENDIX A: SAMPLE PREPARATION}

Unless otherwise specified, all measurements were performed upon an isotopically purified $^{28}$Si crystal obtained from the Avogadro project with 99.995\% $^{28}$Si, $<$\,10$^{14}$ oxygen/cm$^3$ and $<$\,5$\times$10$^{14}$ carbon/cm$^3$ \cite{Becker2010}. An irradiation dose of 320\,kGy was applied to the crystal using 10\,MeV electrons, with intermittent application of cooling dry ice to maintain a relatively low sample temperature during irradiation. To increase hydrogen concentration, the sample was first annealed in boiling water for 24 hours \cite{Ohmura1999}, and then annealed in air from 300\,$^\circ$C to 450\,$^\circ$C in steps of 30 minutes on a hotplate \cite{Lightowlers1994hydrogen}. To reduce laser scatter at the silicon interface, a coarse polish was applied to the larger faces of the sample, followed by a brief etch in a 1:10 HF/HNO$_3$ solution to remove surface strain. 

A second isotopically purified $^{28}$Si crystal, with a higher carbon concentration of 1.5$\times$10$^{15}$\,cm$^{-3}$, was used for the phonon sideband spectra.
The same radiation and annealing treatment was applied to this sample as the first, and the TX PL signal was $\sim$\,3--4 times stronger.
Lastly, a $^\text{nat}$Si sample which was lightly doped with gallium, electron irradiated and then annealed at 500\,$^\circ$C in the year 1995 and stored at room temperature since then was used for the $^\text{nat}$Si PL spectrum. 

\section*{APPENDIX B: METHODS}

\textit{Cryogenics} -- The samples were loosely mounted in a strain-free manner and immersed in liquid helium-4 (LHe$^4$) at temperatures ranging from 1.4\,K to 4.2\,K.  The temperature was set by pumping on the LHe$^4$ bath. 

\textit{Photoluminescence} -- To generate nonresonant photoluminescence (PL), the sample was illuminated with up to 300\,mW of 1047\,nm (beam diameter 2--4\,mm) above-gap excitation. The resulting luminescence from the sample was directed into a Bruker IFS 125 HR Fourier transform infrared (FTIR) spectrometer with a CaF$_2$ beam splitter and liquid nitrogen cooled Ge diode detector, and measured at spectral resolutions ranging from 0.25\,$\upmu$eV to 62\,$\upmu$eV. The apodization method used in shown spectra was Blackman-Harris 3-term, except for the 935.142 meV $^{13}$C peak doublet (Fig.~\ref{fig:photolum}(d)) where boxcar (no apodization) method was used. For the magneto-PL measurements, the sample was centered in a 6\,T superconducting magnet with the field approximately parallel to the [110] orientation. 

For the resonant PL measurements, a single-frequency Toptica DL100 tunable diode laser was amplified by a Thorlabs BOA1017P amplifier to reach powers of \mbox{$\sim\!75$\,mW} (beam diameter 2--4\,mm), then filtered by an Edmund Optics \#87-830 1350\,nm ($\pm$12.5\,nm) bandpass filter and an Iridian Spectral Technologies  DWDM 1329.22 200~GHz 1329\,nm ($\pm$0.5\,nm) bandpass filter before reaching the sample. Both filters were tilted down to shift their respective passbands to the laser frequency. In the optical detection path, two custom 1330\,nm longpass filters (3\,nm cut-on) from Iridian Spectral Technologies were used to filter back-reflected laser light from the sample. The resulting luminescence was directed to the FTIR spectrometer as above.

\textit{Photoluminescence Excitation} --  Photoluminescence excitation (PLE) involves tunable resonant excitation of the TX$_0$ ZPL followed by the optical detection of lower energy photons resulting from the TX phonon sideband. In the case of resonantly driving the 1326\,nm ZPL transition (Fig.~\ref{fig:diagram}(a)), a single-frequency Toptica tunable diode laser DL100 was first amplified by a Thorlabs BOA1017P amplifier to reach powers of $\sim\!100$\,mW (beam diameter 2--4\,mm), then filtered by 1325\,nm ($\pm$25\,nm) band pass filters and directed onto a polished face of the sample. 

In the optical detection path, a Semrock BLP02-1319R-25 1319\,nm longpass laser rejection filter was found to give good rejection of the TX$_0$ pump photons. To remove contributions from a silicon Raman replica at 1426\,nm, a 1375\,nm band pass filter (50\,nm bandwidth) was also in the detection path, giving rise to a spectral PLE detection window as depicted in Fig.~\ref{fig:photolum}(a). The resulting photons were directed to an IDQuantique ID230 high-sensitivity InGaAs photon-counting detector.

PLE was also used for excited state spectroscopy, wherein the Toptica DL100 tunable diode laser was scanned over higher energies and TX$_0$ ZPL photons were detected by replacing the aforementioned detection filters with a 3/4\,m focal length double monochromator set to filter out all light apart from a small window of luminescence around 1326\,nm.

\textit{Optically Detected Magnetic Resonance} -- Unless otherwise specified, the magnetic resonance experiments were performed with the sample in an applied magnetic field of approximately 80\,mT delivered using an iron core electromagnet. The sample was mounted in the PLE setup as described above, and placed within two nested magnetic resonance resonators: a split-ring resonator ($f_\text{res}$\,=\,2.25\,GHz, bandwidth\,=\,10\,MHz) for electron paramagnetic resonance (EPR), and a Helmoltz coil pair ($f_\text{res}$\,=\,4.8\,MHz, bandwidth\,=\,300\,kHz) for nuclear magnetic resonance (NMR). The radiofrequency (RF) signals were generated using signal generators SRS SG384 and SRS SG386, switches ZASWA-2-50DR+, were combined using a power splitter ZB2PD-63-S+ and were amplified to up to 1\,W of power using amplifiers ZHL-16W-43-S+  and ZHL-1-2W+ as needed. For pulse sequencing a Spincore Pulseblaster DDSII-300 was used. 

\FloatBarrier
\bibliographystyle{apsrev4-1}


%

\end{document}